\documentclass[12pt, a4paper]{article}

\pdfoutput=1

\usepackage{amsmath}
\usepackage{amsfonts}
\usepackage{amssymb}
\usepackage{cite}
\usepackage{multicol,multirow}
\usepackage{graphicx, physics, subcaption}
\usepackage{graphics, epsfig}
\usepackage{epsf}
\usepackage{epstopdf}
\usepackage{bm}
\usepackage[usenames,dvipsnames]{color}

\usepackage{booktabs}   
\usepackage{array}

\usepackage{colortbl}
\definecolor{summersky}{cmyk}{0.71,0.33,0,0.14}
\definecolor{flamingo}{cmyk}{0,0.51,0.71,0.14}
\definecolor{rp}{cmyk}{0.2, 1, 0.6, 0}
\definecolor{pacificblue}{cmyk}{0.95,0.3,0, 0.19}
\definecolor{gray60}{cmyk}{0.4,0.4,0,0.8}

\numberwithin{equation}{section}

\usepackage[colorlinks,bookmarks=false, hypertexnames=true]{hyperref}
\hypersetup{
	breaklinks=true,
	pdfstartview={FitH},    
	colorlinks=true,       
	linkcolor=pacificblue,          
	citecolor=flamingo,        
	filecolor=magenta,      
	urlcolor=red,           
	anchorcolor=green,      
	linktocpage=true
}

\newcommand{\nc}{\newcommand}

\nc{\ba}{\begin{eqnarray}}
\nc{\ea}{\end{eqnarray}}

\nc{\calR}{{\cal{R}}}
\nc{\calP}{{\cal{P}}}
\nc{\cN}{ {\cal{N}} }

\nc{\rc}{\textcolor[rgb]{1.00,0.00,0.00}}
\nc{\bc}{\textcolor[rgb]{0.00,0.07,1.00}}
\nc{\gc}{\textcolor[rgb]{0,0.6,0}}

\def\bfk{{\bf k}}

\def\phicr{{\phi_{_{\rm CR}}}}
\def\phicrs{{\phi^2_{_{\rm CR}}}}
\def\phic{{\phi_{{\rm c}}}}
\def\Nf{{{\cal N}_{\mathrm f}}}
\def\nf{{N_{\rm f}}}
\def\Ns{{N_{\star}}}

\begin{document}

\def\thefootnote{\fnsymbol{footnote}}

\begin{center}

{\bf  Models with Non-minimal Coupling in Primordial Universe\\ and Cosmological Observations
}
\\[0.5cm]

{Alireza Talebian$\footnote{talebian@ipm.ir}^{1,\,2}$,
	Hassan Firouzjahi$\footnote{firouz@ipm.ir}^{1}$  and 
	Fereshteh Felegary$\footnote{fereshteh.felegary@ipm.ir}^{1}$ 
	}   
\\[0.5cm]
{\small \textit{ $^{1}$School of Astronomy, Institute for Research in Fundamental Sciences (IPM) \\ P.~O.~Box 19395-5746, Tehran, Iran}}
\\[0.1cm]
{\small \textit{ $^{2}$Department of Physics, Shahid Beheshti University, Tehran 1983969411, Iran}} \\

\end{center}

\vspace{.3cm}
\hrule
\begin{abstract}

A non-minimal coupling between gravity and the inflaton field is a generic contribution in  inflationary cosmology.  In this work, we consider models of inflation with a non-minimal coupling $-\xi\phi^2 R$  and study their observational predictions for the tensor-to-scalar ratio $r$ and the spectral index $n_s$. Unlike the conventional approach of working in the Einstein frame, we perform the analysis in the Jordan frame where  the underlying dynamics, particularly the competition between the potential  force and the coupling-induced friction, are more transparent. For suitable values of  $\xi$, the system exhibits an extended constant-roll  regime, in which this friction counteracts the potential force. Focusing on  monomial potentials $V(\phi) \propto \phi^n$, we find that negative $\xi$ systematically reduces $r$ in the $(n_s, r)$ plane, while the shift in $n_s$ depends on the power $n$: for $n \geq 4$, the spectral index increases with $|\xi|$, whereas for $n < 4$, it decreases. Notably, the quartic model $V(\phi) = \lambda \phi^4/4$ with $\xi \lesssim -0.1$ shows good agreement with the  ACT DR6 data and exhibits a distinct $n_s(\xi)$ dependence compared to other monomial cases. Our results demonstrate that non-minimally coupled models can effectively reconcile Planck and ACT observations, providing a unified framework for interpreting current CMB constraints.
\end{abstract}

\newpage





\section{Introduction}
\label{sec:intro}

The Cosmic Microwave Background (CMB) radiation serves as the most powerful observational probe of the primordial universe. Inflation not only resolves the classical puzzles of the hot big bang cosmology (the horizon, flatness, and monopole problems) but also it provides a quantum mechanical mechanism for the origin of the primordial density perturbations~\cite{Guth:1980zm, Linde:1981mu, Albrecht:1982wi, Mukhanov:1990me, Kodama:1984ziu, Weinberg:2008zzc}. These perturbations imprint themselves on the CMB as temperature and polarization anisotropies, successfully predicting the observed large-scale structure.

For decades, the simple chaotic inflation models with monomial potentials
 \( V(\phi) \propto \phi^n \) (particularly the quadratic model $n=2$) have been the textbook paradigms for inflation. However, the tight constraints from the Planck satellite have dramatically reshaped the landscape of viable inflationary models. Planck 2018 data, combined with BICEP/Keck, placed stringent upper bounds on the tensor-to-scalar ratio (\( r_{0.002} < 0.036 \)) and determined the scalar spectral index with high precision (\( n_s = 0.9651 \pm 0.0044 \)) at 68\% CL~\cite{Planck:2018vyg, Planck:2018jri}. Within the standard Einstein gravity framework, the simple \( m^2 \phi^2 \) model predicts a spectral index (\( n_s \simeq 0.96 \)) that is marginally consistent but a tensor-to-scalar ratio (\( r \gtrsim 0.1 \)) that is ruled out by Planck data. Other models such as the Starobinsky $R^2$ model~\cite{Starobinsky:1980te} and monomial potentials $V(\phi) \propto \phi^n$ with $n=1$ and $n=2/3$ are more compatible with the data~\cite{Planck:2018jri, Dioguardi:2025vci}.

However, the recent data release from the Atacama Cosmology Telescope (ACT) \cite{AtacamaCosmologyTelescope:2025blo, AtacamaCosmologyTelescope:2025nti} has introduced a shift in the preferred value of the scalar spectral index. While the Planck 2018 analysis previously determined $n_s = 0.9651 \pm 0.0044$~\cite{Planck:2018vyg, Planck:2018jri}, the latest joint analysis of Planck and ACT data yields $n_s = 0.9709 \pm 0.0038$. Furthermore, when Planck, ACT, and DESI data are combined (ACT+Planck), the constraint tightens to $n_s = 0.9743 \pm 0.0034$~\cite{AtacamaCosmologyTelescope:2025nti}, deviating from the original Planck-only result by approximately $2\sigma$. Although it is still too early to draw definitive conclusions, this shift has important implications for inflationary model building. Notably, the well-known Starobinsky $R^2$ model, which has been a favoured candidate in the Planck era, is now disfavoured at the $2\sigma$ level by the combined ACT and Planck datasets. Nevertheless, many other inflationary models remain compatible with the latest ACT data release such as setups of non-minimal Coleman-Weinberg inflation~\cite{Marzola:2016xgb, Jarv:2017azx, Racioppi:2018zoy, Kannike:2018zwn, Racioppi:2019jsp, Gialamas:2020snr, Gialamas:2021rpr, Racioppi:2021ynx, Kannike:2023kzt}, non-minimal metric-affine gravity models~\cite{Gialamas:2024uar, Racioppi:2024pno, Racioppi:2024zva, Bostan:2025vkt}, generalized hilltop models~\cite{Lillepalu:2022knx}, polynomial inflation with non-minimal coupling to gravity~\cite{Bezrukov:2007ep, Bezrukov:2014ipa, Kallosh:2013tua, Rubio:2018ogq},  polynomial and hybrid $\alpha$-attractors~\cite{Kallosh:2022feu, Kallosh:2022ggf, Kallosh:2019jnl, Dalianis:2018frf}, supersymmetric hybrid inflation~\cite{Buchmuller:2014epa, Schmitz:2018nhb}, as well as the double inflection point polynomial potential~\cite{Allegrini:2024ooy}. More recently, fractional attractors have been shown to provide a good fit to the ACT data~\cite{Dioguardi:2025vci}, and Palatini linear attractors have also been revisited in the light of the new observations~\cite{Dioguardi:2025mpp}. 

One of the most well-motivated proposals to accommodate the new ACT constraints is the introduction of a non-minimal coupling between the inflaton field and gravity \cite{Kallosh:2025rni, Wang:2025dbj, Yuennan:2025mlg, Yuennan:2026fcn, Ahmed:2025rrg, Gao:2025onc, Gonuguntla:2026rkw}.
Recent studies demonstrate that these models can reduce the tensor-to-scalar ratio while remaining consistent with the latest ACT and Planck measurements. Consequently, 
models of inflation with non-minimal coupling have received significant attentions in a wide range of models, including Higgs inflation, derivative coupling models, induced gravity, and generalized attractor scenarios \cite{Haque:2025uri, Park:2008hz, Haque:2025uga, McDonald:2025tfp}. 
The idea of non-minimal coupling is commonly employed in Higgs inflation, where a sizeable non-minimal coupling flattens the quartic potential in the Einstein frame, allowing the Higgs field to play the role of the inflaton field while remaining consistent with particle physics~\cite{Bezrukov:2007ep, Bezrukov:2009db}. This is well-motivated 
from scalar-tensor theories as they provide a systematic framework to study these effects~\cite{DeFelice:2010aj, Capozziello:2011et, Clifton:2011jh, Kobayashi:2019hrl}. The key advantage of a non-minimal coupling is the additional free parameter $\xi$, which can significantly alter the inflationary predictions in the $(n_s, r)$ plane. In light of the recent ACT data, many studies have revisited non-minimally coupled scenarios, including Higgs inflation with reheating effects~\cite{Liu:2025qca}, chaotic inflation~\cite{Kallosh:2025rni}, hilltop inflation~\cite{Yuennan:2025mlg}, models with non-minimal derivative coupling~\cite{Gao:2025viy}, as well as confronting non-minimal coupling with ACT data ~\cite{Gao:2025onc, Roy:2026vwx}.

In the present work, we consider a single-field inflation setup with a non-minimal coupling, described by the action
\[
S = \int \dd^4 x \sqrt{-g} \left[ \frac{M_{\rm P}^2}{2} f(\phi) R - \frac{1}{2} \partial^\mu \phi \partial_\mu \phi - V(\phi) \right],
\]
where $M_{\rm P}$ is the reduced Planck mass, $R$ is the Ricci scalar, and
\ba
\label{f}
f(\phi) \equiv 1 - \xi \Big(\frac{\phi}{M_{\rm P}}\Big)^2 = 1 + \Big(\frac{\phi}{\phic}\Big)^2, \qquad \phic \equiv \frac{M_{\rm P}}{\sqrt{-\xi}},
\ea
with $\xi$ being a constant parameter, in which, as we shall see below, $\xi<0$ to fit the ACT data.  The potential $V(\phi)$ is kept general with monomial form $V \propto \phi^n$. However, we pay particular attention to the special cases of $n=2$ (the quadratic potential $m^2\phi^2/2$) and $n=4$ (the quartic potential $\lambda\phi^4/4$). Indeed, the setups with a non-minimal coupling for $n=2$ and $n=4$ are vastly studied in the context of quantum field theory in curved spacetime; for a review, see \cite{Parker:2009uva, Birrell:1982ix}, in which the parameter $\xi$ is interpreted as the conformal coupling.  The case $n=4$ is particularly interesting, in which 
$\lambda$ is dimensionless, and the theory is classically conformal invariant when $\xi=1/6$.  However, it is a well-known result that conformal invariance is broken under quantum loop corrections \cite{Onemli:2002hr, Brunier:2004sb, Prokopec:2008gw, Kahya:2009sz, Firouzjahi:2023wbe}.

It is important to note that most previous analyses in the literature involving setups with a non-minimal coupling have been performed in the Einstein frame, where a conformal transformation is used to recast the action into the canonical form of general relativity. In the Einstein frame, the physical effects of the coupling parameter $\xi$  are absorbed into a redefinition of the metric, the inflaton field, and the scalar potential. However, the physical interpretation of the fields and quantities is more transparent in the original Jordan frame. In this frame, the non-minimal coupling directly modifies the effective gravitational coupling and the background evolution of the inflaton, and its effects can be explicitly tracked in the final expressions for the observables $n_s$ and $r$. Therefore, in this work we perform all analysis in the  Jordan frame to gain a clear understanding of the underlying physics. We aim to derive analytical expressions for $n_s$ and $r$ and confront them with the latest Planck and ACT data, thereby assessing the viability of these non-minimally coupled models in light of the new observations. Having said this, the physical results obtained in both frames are equivalent; it is only the question of mathematical simplicity and interpretation of the results that we believe are more transparent in the original Jordan frame, where the potential has a simple form. 

Another important point to take into account is that we work with the monomial potentials $V(\phi) \propto \phi^n$
in which $\phi$ rolls to its global minimum $\phi=0$ at the end of inflation. As a result, the correction in the effective gravitational coupling vanishes 
when the field settles down to its global minimum and  $M_P$ will be the actual gravitational coupling after inflation and reheating. 
Since the cosmological observables are measured at the end of inflation, 
there will be no complication in dealing with a time-dependent gravitational coupling.
This was the main reason to work in the Einstein frame in alternative approaches to get rid of a time-varying gravitational coupling (i.e., time-dependent Newton constant).

The rest of the paper is organized as follows. In Section~\ref{sec:background}, we study the background dynamics of the scalar field in the Jordan frame, analyzing both the slow-roll (SR) and constant-roll (CR) regimes. In Section~\ref{sec:perturbations}, we compute the second-order actions for the scalar and tensor perturbations and obtain analytical expressions for the inflationary observables such as  $n_s$ and $r$.  In Section~\ref{sec:CMB}, we confront the model predictions with the latest CMB data from Planck 2018 and ACT DR6, placing constraints on the model parameters.  Finally, Section~\ref{sec:conclusions} summarizes our findings and discusses future directions.

\section{Background Dynamics}
\label{sec:background}

In this section, we study the background dynamics in the presence of non-minimal coupling. 

The non-minimal coupling modifies the field equations. The modified Klein-Gordon  (KG) equation reads,
\ba
\label{KG}
\Box \phi - \xi R \phi - V_{, \phi} = 0.
\ea
On the other hand,  the Einstein field equation, in the presence of $\xi$ and the field-dependent gravitational coupling $f(\phi)$ defined in Eq. (\ref{f}),  is written as follows \cite{Kofman:2007tr, Firouzjahi:2023wbe},
\ba
\label{Einstein}
M_{\rm P}^2 f(\phi) G_{\mu \nu}= \partial _\mu \phi \partial _\nu \phi -
\frac{1}{2}g_{\mu \nu } {\partial ^\alpha }\phi {\partial _\alpha }\phi
-\xi \big[   \nabla_\mu \nabla_\nu (\phi^2) - \nabla^\rho \nabla_\rho (\phi^2) g_{\mu \nu}   \big] - g_{\mu \nu} V(\phi)  \, .
\ea

This yields to the following background equations \cite{Kofman:2007tr}, 
\ba
\label{H-eq}
3 M_{\rm P}^2 f(\phi)  H^2= \frac{\dot \phi^2}{2} + 6 \xi H \phi \dot \phi + V(\phi),
\ea
and
\ba
\label{phi-eq}
\ddot \phi + 3 H \dot \phi + 6 \xi (\dot H + 2 H^2) \phi + V_{,\phi} =0 \, ,
\ea
where $H$ is the Hubble expansion rate and a dot indicates the derivative with respect to the cosmic time $t$.

Eq. (\ref{phi-eq}) suggests that the effective mass squared of the inflaton receives the additional contribution  $12\xi H^2$ from the non-minimal coupling.  This is understood from the starting Lagrangian in which $\xi R \simeq 12 \xi H^2 $ appears as a mass term. Correspondingly, to support a period of SR 
inflation, we expect that $\xi$ to be small, roughly speaking 
$\xi \sim m^2/H^2 \ll 1$. Furthermore, a negative (positive) $\xi$ contributes negatively (positively) to the effective mass. Therefore, for $\xi<0$ the rolling towards the bottom of the potential is slowed down, causing a prolonged period of inflation compared to the case $\xi>0$ where the opposite effect occurs.  An interesting effect observed from this equation is that the force induced by this effective mass term can balance the driving force from the potential, yielding a  period of constant-roll (CR) inflation. For the monomial potential $V(\phi) \propto \phi^n$, a CR phase happens when $\xi<0$. This effect is easily observed here in the Jordan frame, while it is not easy to track this effect in the Einstein frame, in which the effects of $\xi$ are translated into a modification of the potential and the redefinition of the inflaton field.

Combining Eqs. (\ref{H-eq}) and (\ref{phi-eq}), we find the following equation for the evolution of the Hubble expansion rate,
\ba
\label{Hdot}
- 2 M_{\rm P}^2 f(\phi) \dot H = (1+ M_{\rm P}^2 f_{, \phi \phi} ) \dot \phi^2 + M_{\rm P}^2 f_{, \phi} (\ddot \phi - H \dot \phi) \,.
\ea

Using the number of e-folds $\dd N= H \dd t$ as the clock, the background equations (\ref{H-eq}) and (\ref{phi-eq}) can be written as follows, 
\ba
\label{H-eq-new}
3 M_{\rm P}^2 H^2 f(\phi) = \frac{H^2 \phi^{\prime^2}}{2} + 6 \xi H^2 \phi \phi^{\prime} +  V(\phi),
\ea
and
\ba
\label{phi-eq-new}
\phi^{\prime\prime} + (3-\epsilon)\phi^{\prime} + 6 \xi (2-\epsilon)\phi +H^{-2}~ V_{,\phi} =0 \, ,
\ea
where a prime indicates  the derivative with respect to the number of e-folds $N$, and $\epsilon$ is the first slow-roll (SR) parameter, which, along with the second SR parameter $\eta$, is defined via, 
\ba
\epsilon\equiv  -\frac{\dot H}{H^2} , \quad \quad
\eta\equiv \frac{\dot \epsilon}{H \epsilon} \,.
\ea
In the SR regime, the Hubble expansion rate  from Eq. (\ref{H-eq-new}) can be approximated as
\ba \label{H2-eq1} H^2  \simeq  \frac{V(\phi)}{3 \big( M_{\rm P}^2 - \xi \phi^2 - 2 \xi \phi \phi'   \big)} \, , \ea
where we have safely ignored the kinetic energy involving $\phi'^2$.

\begin{figure}[t]
	\vspace{-0.5cm}
	\begin{center}
		\includegraphics[scale=1.2]{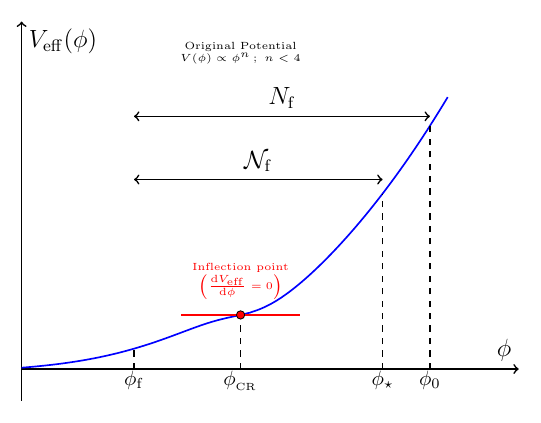}
	\end{center}
	\caption{ {Schematic illustration of the effective potential $V_{\mathrm{eff}}(\phi)$ defined in  Eq. \eqref{Veff}. Here $\phi_0$ denotes the initial value of the scalar field, $\phi_\star$ is the field value when the CMB mode leaves the horizon, $\phicr$  is the critical field value given by Eq. \eqref{phiCR}, where the system enters a CR phase, and $\phi_f$ defined in  Eq. \eqref{phi_f}, is the value of inflaton at the end of inflation. 
$\nf$ denotes the total duration of inflation from $\phi_{0}$ to $\phi_{f}$
while $\Nf$ represents the number of e-folds from the point of CMB 
horizon exit  to the end of inflation.}}
	\label{fig:V_eff}
\end{figure}
As mentioned before,  a CR phase can arise from the interplay of the two competing terms, the last two terms
in (\ref{phi-eq}). This corresponds to the following equation,
\ba
\label{Veff}
V_{{\rm eff},\phi} \equiv 6 \xi H^2 (2-\epsilon)  \phi + V_{,\phi} \simeq 0. 
\ea
The above condition may hold for a limited range of $\phi$, allowing for a period of CR inflation. In Fig. \ref{fig:V_eff}, we have illustrated the effective potential $V_{\rm eff}$ and represented the field values during the inflationary era schematically. Considering Eq. \eqref{H2-eq1} in the SR limit and in the limit of small coupling $|\xi| \ll 1$ in which the dominant contribution to the Hubble rate comes from the scalar potential, Eq. (\ref{Veff}) determines an ``{inflection point}'' 
for the monomial potential $V(\phi)\propto \phi^{n}$ whose value can be estimated as,
\begin{align}
	\label{phiCR}
	\phicr \simeq \sqrt{\frac{n}{4-n-2\epsilon}}\phic \,,  
\end{align}
in which $\phic$ is defined in Eq. (\ref{f}). Note that for $n \ge 4-2\epsilon$  there is no inflection point in our case (i.e. $\xi <0$)
and the field does not experience a CR regime. 

We define the CR parameter $\beta$ via \cite{Motohashi:2014ppa, Motohashi:2017aob, Motohashi:2019tyj},
\begin{align}
	\label{beta}
	\beta \equiv \frac{\phi''}{\phi'} \, ,
	\end{align}
in which a CR phase corresponds to a period during inflation in which $\beta$ is nearly constant. The case of ultra slow-roll (USR) \cite{Kinney:2005vj, Namjoo:2012aa, Martin:2012pe} corresponds to the special situation where $\beta=-3$. {Defining $\beta$ according to \eqref{beta} and making use of \eqref{H2-eq1}, the KG equation \eqref{phi-eq-new} can be rewritten in the following compact form,
\ba
	\label{phi-eq2}
	(\phi^2)' - \dfrac{6 \xi(4-n- 2 \epsilon)}{3(2 n \xi -1)+\epsilon-\beta} \left( \phi^2 - \phicrs \right) \simeq 0 \, .
\ea 
Under the assumption that $\epsilon$ and $\beta$ vary slowly and can be treated as constants over the period of interest, this equation admits a straightforward analytical solution.} 

To present the numerical results,  we consider the specific models {listed} in Table \ref{table:nonminimal} and compare them {with the corresponding minimal cases presented} in Table \ref{table:minimal}. These values have been selected to satisfy the COBE normalization and to obtain $\Nf = 60$ e-foldings to be able to compare the minimal and non-minimal models in a single plot  as presented in  {Figures}. \ref{fig:quartic}, \ref{fig:phi_onethird-SR}, \ref{fig:phi_onethird-CR}, and \ref{fig:quadratic}. The models considered are as follows, 
	\begin{itemize}
		\item \textbf{Model NI:} Quartic potential $V(\phi)=\frac{1}{4}\lambda  \phi^4$ in SR regime (Fig. \ref{fig:quartic}).
		\item \textbf{Model NII:} Fractional power-law $V(\phi)=\lambda M_{\rm P}^{11/3} \phi^{1/3}$ in SR regime (Fig. \ref{fig:phi_onethird-SR}) and CR regime (Fig. \ref{fig:phi_onethird-CR}).
		\item \textbf{Model NIII:} Quadratic potential $V(\phi)=\frac{1}{2} m^2  \phi^2$ in CR regime (Fig. \ref{fig:quadratic}).
	\end{itemize}
	
In the following subsections, we obtain analytical solutions for the background dynamics in the SR and CR regimes and compare them with the full numerical results. 
	
	\vspace{0.5cm}
	
	\begin{table}[h!]
		\centering
		\caption{Non-minimal models}
		\begin{tabular}{l c c c c l l}
			\toprule
			\textbf{Model} & $\phi_0/M_{\rm P}$ & $\xi$ & $\phi_c/M_{\rm P}$ & $\phi_{\rm cr}/M_{\rm P}$ & \textbf{Potential} & \textbf{Parameter} \\
			\midrule
			$\mathrm{NI}$  & 17.3   & -0.1   & 3.16  & --    & $V \propto \phi^4$ (Quartic)    & $\lambda=1.5 \times 10^{-10}$ \\
			$\mathrm{NII-SR}$   & 5.65   & -0.001 & 31.6  & 9.53  & $V \propto \phi^{1/3}$           & $\lambda=2.1 \times 10^{-10}$  \\
			$\mathrm{NII-CR}$ & 2.9932 & -0.01  & 10    & 3.01  & $V \propto \phi^{1/3}$           & $\lambda=4.85 \times 10^{-13}$ \\
			$\mathrm{NIII}$  & 4.9993 & -0.04  & 5     & 5     & $V \propto \phi^2$ (Quadratic)  & $m=2.5 \times 10^{-8} M_{\rm P}$  \\
			\bottomrule
		\end{tabular}
		\label{table:nonminimal}
		\vspace{0.5cm}
	\end{table}

	\begin{table}[h!]
		\centering
		\caption{Minimal models ($\xi=0$)}
		\begin{tabular}{l c l l}
			\toprule
			\textbf{Models} & $\phi_0/M_{\rm P}$ & \textbf{Potential} & \textbf{Parameter} \\
			\midrule
			$\mathrm{MI}$     & 21.9   &  $V \propto \phi^4$ (Quartic)   & $\lambda=1.45 \times 10^{-13}$ \\
			$\mathrm{MII}$ & 6.275  & $V \propto \phi^{1/3}$         & $\lambda=3.8 \times 10^{-10}$ \\
			$\mathrm{MIII}$     & 15.4   & $V \propto \phi^2$ (Quadratic) & $m=6.3 \times 10^{-6} M_{\rm P}$  \\
			\bottomrule
		\end{tabular}
		\label{table:minimal}
		\vspace{0.5cm}
	\end{table}


\subsection{Slow-Roll Inflation for $\bm{V(\phi)\propto \phi^{n}}$}

In this subsection, we solve the background equations \eqref{H-eq-new} and \eqref{phi-eq-new} in the SR regime. To perform the analytic calculations, 
we consider a monomial potential $V \propto \phi^n$.  One can consider the more general case of non-monomial potential, such as the symmetry breaking (i.e., Higgs-like) potential as well.

In the SR regime, one can neglect the acceleration term $\phi''$ in the Klein-Gordon equation \eqref{KG} or equivalently set $\beta=0$ in \eqref{phi-eq2} to obtain
\ba
\label{phi-eq-SR}
(\phi^2)' - \mu \left( \phi^2 - \phicrs \right) \simeq 0 \, ,
\ea
in which the dimensionless constant $\mu$ is defined via, 
\begin{align}
	\mu \equiv \dfrac{2 \xi(4-n- 2 \epsilon)}{2 n \xi -1}\, .
\end{align}
Here we have discarded the term $\epsilon$ that appears in the denominator of \eqref{phi-eq2} at the leading order in the SR expansion.  However, we keep the term containing the combination   $\epsilon \xi$ in the numerator. This is because  for the special case  $n=4$ the two leading terms in the numerator cancel each other so we keep the subleading term involving  $\epsilon$. The fact that the coefficient of $\xi \phi^2$ is cancelled to leading order for $n=4$ is a unique feature of $\lambda \phi^4$ setup where the model is classically conformal invariant. 

Solving Eq. (\ref{phi-eq-SR}) yields,
\ba
\label{phi-sol-SR}
\phi(N)^2 \simeq \phicrs + \big( \phi_0^2 - \phicrs \big) \, e^{\mu N}
	 \,; 
\hspace{1.5cm}
\text{(SR phase)}
\ea
in which $\phi_0$ is the initial value of the field at the start of inflation, which we set to $N=0$, so $\phi_0= \phi(N=0)$.

Inflation proceeds until the SR conditions are violated. As a measure of when this happens, let us look at the first SR parameter $\epsilon$. Using Eq. (\ref{H2-eq1}) for 
$H$, we obtain,
\ba
\label{epsilon_H}
\epsilon = - \frac{H'}{H} \simeq - Q \frac{\phi'}{\phi}
\,;
\hspace{1cm}
Q \equiv \frac{n}{2}-\frac{1}{1+(\frac{\phi_{c}}{\phi})^2}
 \,.
\ea
Considering $\epsilon \simeq 1$ at the end of inflation and using our analytic formula Eq. (\ref{phi-sol-SR}) yields the duration of inflation as follows, 
\ba
\label{N_f_n}
\nf \simeq \frac{1}{\mu} \ln \Bigg[    \frac{ 2\phicrs}{\left(2+\mu \, Q_{\rm f}\right) ( \phicrs - \phi_0^2)}
\Bigg] \, ,
\ea
in which $Q_{\rm f}$ is given by \eqref{epsilon_H} for $\phi = \phi_{\rm f}$, the field value at the end of inflation. {Note that we also use the notation $\Nf = \nf - \Ns$ to denote the number of e-folds between when the CMB mode leaves the horizon ($\Ns$) and the time of end of inflation ($\nf$) as shown in Fig. \ref{fig:V_eff}. 

Plugging Eq. \eqref{N_f_n} in Eq. (\ref{phi-sol-SR}), one finds $\phi_{\rm f}$ as follows
\begin{align}
	\label{phi_f}
	\frac{\phi_{\rm f}}{M_{\rm P}} \simeq \frac{n}{\sqrt{2}} - \frac{n-10}{4\sqrt{2}}n^2 \xi + {\cal O}(\xi^2) \,.
\end{align}
The first term above gives the value of $\phi_{\rm f}$  in the models with no coupling. As seen, the effect of the coupling $\xi < 0$ is to reduce $\phi_{\rm f}$ for $n<10$ compared to models with no coupling. 


\begin{figure}[t]
	\vspace{-0.5cm}
	\begin{center}
		\includegraphics[scale=0.45]{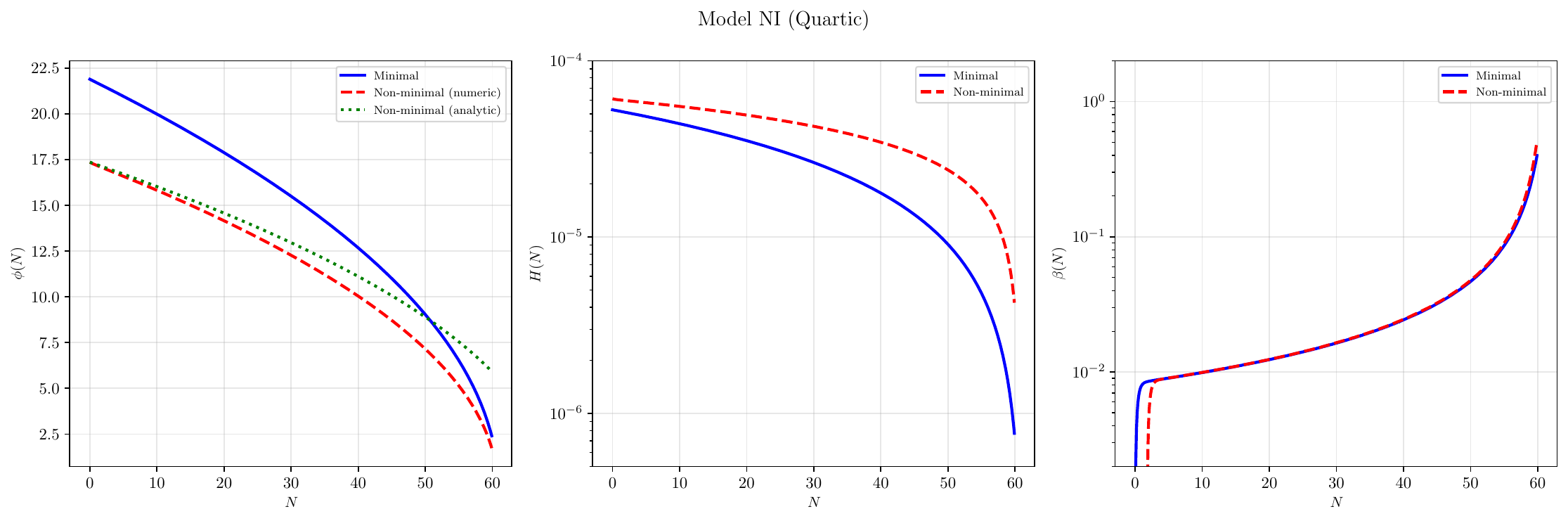}
	\end{center}
	\caption{The evolution of  $\phi(N)$,  $H(N)$ and the CR parameter  $\beta$  (defined in Eq. \eqref{beta}) for the quartic potential $V(\phi)=\frac{1}{4}\lambda\phi^4$.
The solid blue curves correspond to the minimal model (Model MI in Table \ref{table:minimal}), while the dashed red curves show the non-minimal case (Model NI in Table \ref{table:nonminimal}). The dotted green curve in the left panel represents the analytical solution  Eq. \eqref{phi-sol-SR}  
for $\phi(N)$. Deviations between the analytical and full numerical results  appear where the higher-order corrections  ${\cal O}(\epsilon\, N)$ are ignored in the  analytical solution Eq. \eqref{phi_alpha_4_appr}.
	}
	\label{fig:quartic}
\end{figure}

In Figs. \ref{fig:quartic} and \ref{fig:phi_onethird-SR}, we have plotted the background evolutions for the scalar field $\phi(N)$, the Hubble expansion rate $H(N)$, and the CR parameter  $\beta$  (defined in Eq. \eqref{beta})
for Models NI and NII-SR listed in Table \ref{table:nonminimal} and compared them with the corresponding minimal models MI and MII in   
Table \ref{table:minimal}. The analytic solution for $n=1/3$ is in excellent agreement with the full numerical solution  while this agreement is less pronounced for $n=4$. This is due to the fact that for the quartic model, the analytic solution \eqref{phi-sol-SR} is obtained by expanding in powers of $\epsilon N$. Setting $n=4$ in this expansion, we find
	\begin{align}
		\label{phi_alpha_4_appr}
		\phi(N) \simeq \sqrt{\phi_0^2 + \frac{8 \xi N}{1- 8\xi} \phi_c^2} + \mathcal{O}(\epsilon N)^2 \quad \quad \quad (n=4) \, .
	\end{align}
	Since $\epsilon$ becomes $\mathcal{O}(1)$ near the end of inflation (with $N\sim 60$), the expansion parameter $\epsilon N$ is no longer small, and the analytic approximation inevitably deviates from the full numerical solution.

We have considered parameters that generate approximately $\Nf \simeq 60$ e-folds of inflation.  As shown in the left panel of Fig. \ref{fig:quartic} the initial value of the field $\phi_0$ for the case of non-minimal coupling is less than the initial value of the field in the absence of coupling. This indicates that the conformal coupling $\xi$ causes the rolling of the inflaton to slow down, causing a prolonged period of inflation compared to the case $\xi=0$ with the same initial value $\phi_0$.  The same effect can be seen in the middle panel of Figure~\ref{fig:quartic}, in which $H$ is reduced compared to the case of minimal coupling. Overall, the conclusion is   that a negative non-minimal coupling $\xi$ reduces the rate of the evolution of the background quantities.


\begin{figure}[t]
	\vspace{-0.5cm}
	\begin{center}
		\includegraphics[scale=0.45]{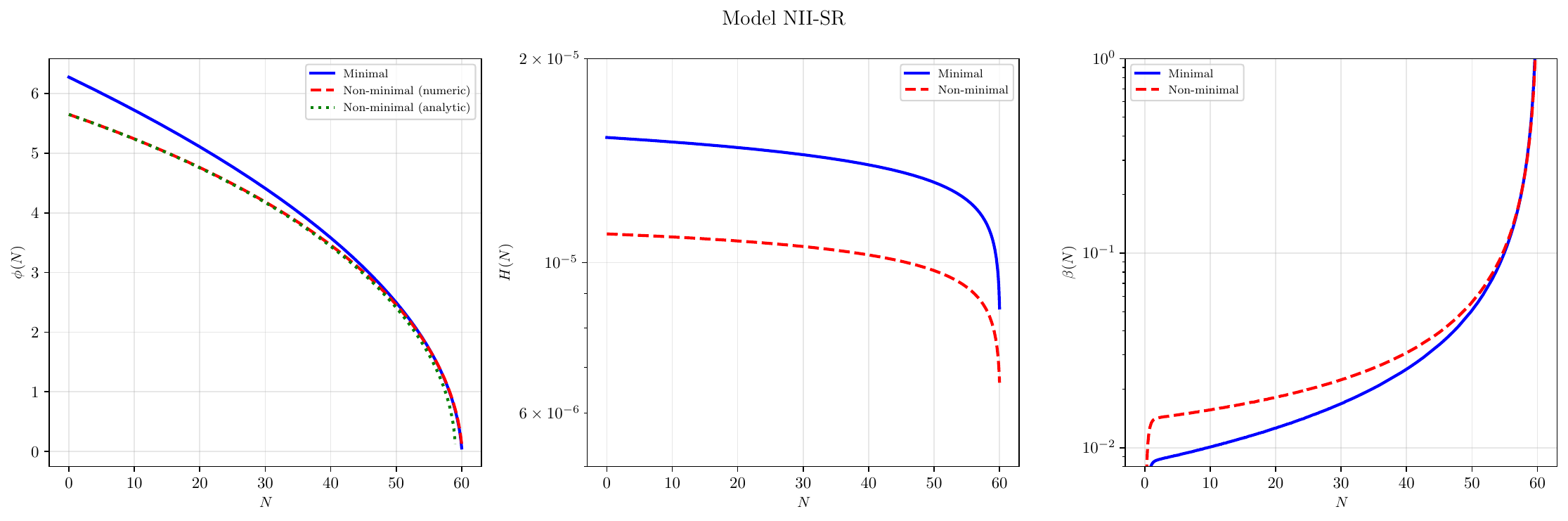}
	\end{center}
	\caption{ The same as in Fig. \ref{fig:quartic} but for  the fractional power-law potential $V(\phi)=\lambda M_{\rm Pl}^{11/3}\phi^{1/3}$. The solid blue curves correspond to the minimal model (Model MII in Table \ref{table:minimal}), whereas the dashed red curves represent the non-minimal case (Model NII-SR in Table \ref{table:nonminimal}). The dotted green curve in the left panel shows the analytical solution given by Eq. \eqref{phi-sol-SR}, which is in excellent agreement with the full numerical solution throughout the SR evolution. 
}
\label{fig:phi_onethird-SR}
\end{figure}


\subsection{Constant-Roll Inflation for $\bm{V(\phi)\propto \phi^{n}}$}

As we mentioned before, working in the Jordan frame allows one to easily understand the existence of a period of the CR phase in some regions of the parameter space.

Consider  the monomial potential $V(\phi)\propto \phi^{n}$. In this case, Eq. (\ref{H2-eq1}) is still valid as there is no competition in the Hubble expansion rate either in the SR or CR stage. Now, keeping the acceleration term in the KG equation and discarding the SR suppressed term as before, we end up with the following equation for the evolution of $\phi(N)$, 
\ba
\label{phiprimnew}
\phi''  - 3(2n\xi-1)\phi' + \frac{3\xi(4-n-2\epsilon)}{\phi} 
\big( \phi^2 - \phicrs \big)  \simeq 0 \, .
\ea
This equation demonstrates that the evolution of $\phi(N)$ is governed by the competition among the acceleration term, the Hubble friction term, and an effective force term induced by the scalar potential and the non-minimal coupling. In the neighbourhood of the critical field value $\phicr$, the effective force is weakened.  As a consequence, the acceleration term becomes non-negligible, and the dynamics temporarily departs from the SR regime, entering a phase characterized by an approximate CR evolution.

Around the critical field value $\phicr$, the effective potential features an approximately stationary stage. The resulting suppression of the effective force restricts $\phi$ to a small neighbourhood of $\phicr$, where $\abs{\phi - \phicr} \ll \phicr$. Therefore one can approximate $\phi^2- \phicrs \simeq 2 \phicr (\phi- \phicr)$ and  solve Eq. (\ref{phiprimnew}),  obtaining the following approximate solution for $\phi(N)$,
\begin{align}
	\label{phi_CR}
	\phi(N) \simeq \phicr + 
	 \frac{\phi_0'}{\kappa_1- \kappa_2} \Big[ e^{\kappa_1 (N-N_{_{\rm CR}})} - e^{\kappa_2 (N-N_{_{\rm CR}})} \Big]
	 \,; 
	 \hspace{1.5cm}
	 \text{(CR phase)}
\end{align} 
in which $\phi(N=N_{_{\rm CR}})=\phicr$, $\phi_0'=\phi'(N=N_{_{\rm CR}})$, and
\ba
\label{kappa12}
\kappa_{1, 2} \equiv 3 n \xi  -\frac{3}{2} \pm \frac{1}{2}\sqrt{9-12\xi(n+8)+36n^2\xi^2}  \, .
\ea
Note that  $\kappa_1 >0$ while $\kappa_2<-3. $ Correspondingly, one can neglect the decaying contribution in solution (\ref{phi_CR}) involving  $e^{\kappa_2 N}$. 

Neglecting the decaying part in $\phi(N)$, we obtain the CR parameter $\beta$ (defined in \eqref{beta}) as follows
\ba
\label{beta_kappa}
\beta\simeq \kappa_1 &=& 3 n \xi  -\frac{3}{2} + \frac{1}{2}\sqrt{9-12\xi(n+8)+36n^2\xi^2} \,  \\
&\simeq& 2 \xi (n-4) + {\cal O}( \xi^2) \, .
\ea
The above equation indicates that a CR phase does
not materialize for the quartic potential $n=4$. This can be seen in the right panel of Fig. \ref{fig:quartic} in which $\beta$ is obtained to be at the order of SR parameters with no difference between the cases of the minimal and non-minimal coupling.


\begin{figure}[t]
	\vspace{-0.5cm}
	\begin{center}
		\includegraphics[scale=0.45]{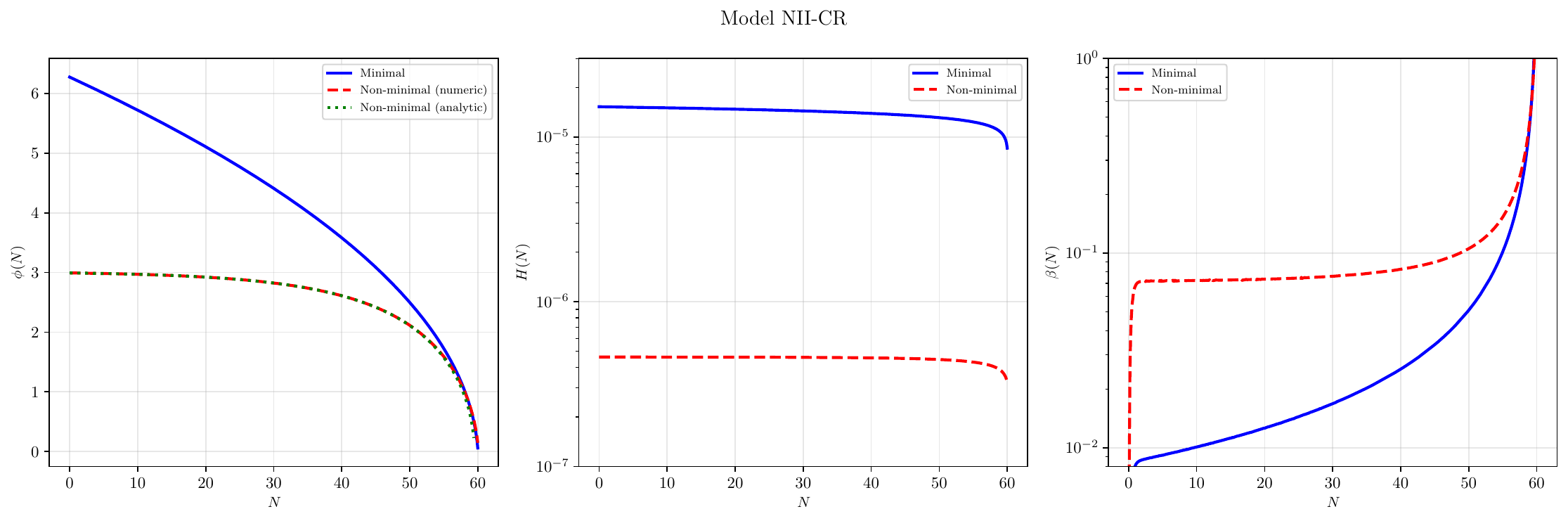}
	\end{center}
	\caption{The same as in Fig. \ref{fig:phi_onethird-SR}, $V(\phi)=\lambda M_{\rm Pl}^{11/3}\phi^{1/3}$, but for the CR phase (Model NII-CR).  The solid blue curves correspond to the minimal model (Model MII in Table \ref{table:minimal}), while the dashed red curves represent the non-minimal model (Model NII-CR in Table \ref{table:nonminimal}). Substituting $\xi= -0.01$  into 
Eq. \eqref{beta_kappa}  yields $\beta \simeq 0.07$, in relatively good agreement with the numerical result shown in the right panel.  The dotted green curve in the left panel shows the analytical solution \eqref{phi-sol-SR}, obtained by replacing the parameter $\mu$ according to Eq. \eqref{phi-eq2} with $\beta=0.07$. The analytic solution is in very good agreement with the numerical evolution of $\phi(N)$.}
	\label{fig:phi_onethird-CR}
\end{figure}
\begin{figure}[t]
	\vspace{-0.5cm}
	\begin{center}
		\includegraphics[scale=0.45]{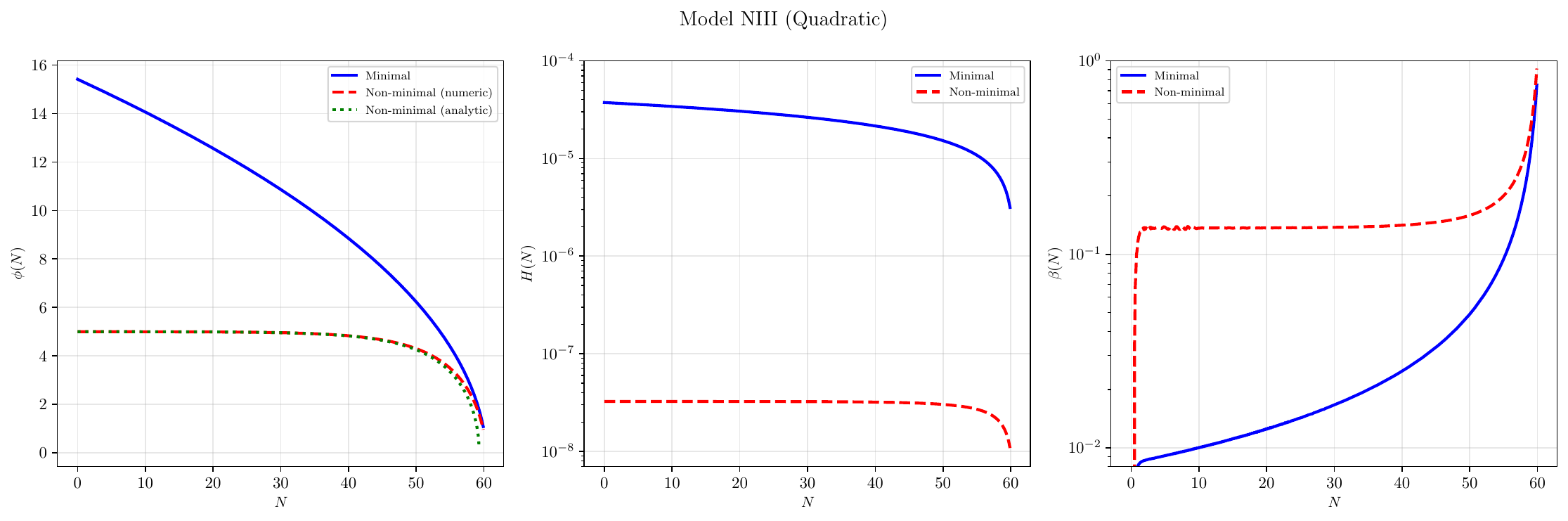}
	\end{center}
	\caption{ Background dynamics in CR phase for the quadratic potential $V(\phi)=\frac{1}{2}m^2\phi^2$ (Model NIII). 
The solid blue curves represent the minimal case (Model MIII in Table \ref{table:minimal}), while the dashed red curves correspond to the non-minimal model (Model NIII in Table \ref{table:nonminimal}). The dotted green curve in the left panel shows the analytic solution given by Eq. \eqref{phi_CR}, which is in good agreement with the full numerical solution. For $\xi= -0.04$, Eq. \eqref{beta_kappa} yields $\beta \simeq 0.133$, in relatively good agreement with the numerical result shown in the right panel.}
	\label{fig:quadratic}
\end{figure}

Figure \ref{fig:phi_onethird-CR} illustrates the background evolution for the chaotic model $V \propto \phi^{1/3}$ in the CR regime when the initial field value is chosen very close to the critical value, $\phi_{0} \simeq \phi_{CR}$. In this limit, $\phi$ rolls much more slowly and 
the parameter $\beta$  becomes nearly constant during the evolution. 
As a result,  an extended period of CR phase exists. Also we see that 
the analytic solution  Eq. \eqref{phi-sol-SR}, after substituting the parameter $\mu$ from Eq. \eqref{phi-eq2}, is in excellent agreement with the full numerical solution.  

Similarly, in Fig. \ref{fig:quadratic}, we have plotted the background solutions in the CR phase for the chaotic model $V= m^2 \phi^2/2$. 
The numerical parameters  used to generate this plot are presented in Table \ref{table:nonminimal}.   In particular, note that the initial value 
$\phi_0$ is significantly smaller than the standard case of chaotic inflation with  $\xi=0$, which requires a large super-Planckian field value to obtain the same number of e-folds. Also note that the mass $m$ is about two orders of magnitude smaller than its value in the minimal setup which requires $m\sim 10^{-6} M_P$ to fit the COBE normalization.

\section{Perturbations}
\label{sec:perturbations}

After studying the background dynamics, we now study the perturbations to investigate the effects of the non-minimal coupling on cosmological observables such as the spectral index $n_s$ and the tensor-to-scalar ratio $r$. 

Starting with the usual ADM formalism, we decompose the metric perturbations as follows, 
\ba
{\rm d} s^2= - N^2 {\rm d} t^2 +h_{ij} ( {\rm d}x^i+ N^i {\rm d} t) ({\rm d} x^j+ N^j {\rm d} t) \, , 
\ea
in which $N$ and $N^i$ are the lapse and the shift functions respectively while $h_{ij}$ is the metric on the constant $t$ hypersurface.

The Ricci scalar $R$ is decomposed as \cite{Poisson:2009pwt, Dyer:2008hb},  
\ba
\label{GC}
R= {^{(3)}}R+ \frac{1}{N^2} \big( E^{i j}E_{ij} - E^2 \big)
-2 \nabla_\alpha \big( n^\beta \nabla_\beta n^\alpha - \frac{n^\alpha}{N} E)\, , 
\ea
in which $E_{ij}$ is the extrinsic curvature,
\ba
E_{ij} = \frac{1}{2} \big( \dot h_{ij} - \nabla_i N_j - \nabla_j N_i \big)\, ,
\ea
with $\nabla_i$ being the covariant derivative with respect to $h_{ij}$, 
$E^{ij}= h^{im} h^{jn} E_{mn}$ and $E= E^i_i$. Here $n^{\alpha}$ is the unit normal vector to the surface of constant 
$t$ with components, 
\ba
n^0= \frac{1}{N}, \quad \quad 
n^i= -\frac{N^i}{N} \, .
\ea

A crucial point to observe is that the last term in Eq. (\ref{GC}) is a covariant derivative, so its effects disappear when calculating the 
action in ADM formalism in standard GR \cite{Maldacena:2002vr} where $f(\phi)=1$. However, in the presence of the non-minimal coupling $f(\phi)$, this boundary term plays important roles and can not be discarded \cite{Dyer:2008hb}. Otherwise, this leads to the wrong equation of motion.

We work in unitary gauge in which $\phi= \phi(t)$ and $\delta \phi=0$. Decomposing the metric perturbations into the scalar and tensor perturbations, we write
 \cite{Maldacena:2002vr},
\ba
h_{ij} = a(t)^2 e^{2 \zeta} \, \big( \delta_{ij} + \gamma_{ij} \big) \, ,
\ea
in which $\zeta$ is identified as the scalar curvature perturbation. On the other hand, $\gamma_{ij}$ represents the tensor perturbations subject to the following transverse and traceless conditions,
\ba
\gamma_{ii}=0\, \quad \quad \partial_i \gamma_{ij} =0 \, ,
\ea
in which the indices on $\gamma_{ij}$ are raised and lowered by $\delta_{ij}$. 
\subsection{Scalar Perturbations}

Let us first consider the scalar perturbations. Using the decomposition (\ref{GC}), the action in comoving gauge  takes the following form,
\ba
\label{action-com}
S= \frac{1}{2} \int {\rm d}^4 x \, a^3 e^{3 \zeta} \Big[ M_{\rm P}^2  f(\phi) \Big(N  {^{(3)}}R + 
\frac{1}{N} ( E_{ij} E^{ij} - E^2) \Big) - 2 M_{\rm P}^2 f_{,\phi} \frac{E}{N} \dot \phi + 
 \frac{\dot \phi^2}{N}  - 2 N V(\phi) \Big]  ,
\ea
in which  $f(\phi)$ has been defined in \eqref{f}. In particular, note the non-trivial contribution involving the term $f_{,\phi}$ in the action above 
which originates from the total derivative term in Eq. (\ref{GC}). This term does not exist in standard GR in which $f(\phi)=1$.

Let us define the scalar perturbations in lapse  and shift functions as follows \cite{Maldacena:2002vr},
\ba
N\equiv 1+ N_1(t, x, y, z) \, , \quad \quad
N^i\equiv \partial_i\psi(t, x, y, z) \, ,
\ea  
which yields, 
\ba
E_{ij}= a(t)^2 e^{2 \zeta} \big[ (H+ \dot \zeta) \delta_{ij} - \partial_i \partial_j \psi
- \delta_{ij} \partial \psi\cdot \partial\zeta   \big] \, , 
\quad 
E= 3(H+ \dot \zeta) - \partial^2 \psi - 3 \partial \psi \cdot \partial \zeta \, ,
\ea
and
\ba
{^{(3)}}R= -e^{-2 \zeta} a(t)^{-2} \big(  4 \partial^2 \zeta + 2 ( \partial \zeta)^2
\big) \, .
\ea
Correspondingly, to second order, we obtain, 
\ba
E_{ij} E^{ij} - E^2 \simeq -6 (H+ \dot \zeta)^2 + (\partial_i \partial_j \psi)^2 
-(\partial^2 \psi)^2 + 4 (H+ \dot \zeta) \partial^2 \psi + 12 H \partial \psi \cdot \partial \zeta \, .
 \ea

Plugging the above results into the action (\ref{action-com}), and varying the action  with respect to $N_1$ and $\psi$ yields  the following constraint equations respectively,
\ba
\label{pert-eq1} 
M_{\rm P}^2  \Big(- f(\phi) ( E_{ij} E^{ij} -  E^2) +  2 f_{,\phi} E \dot \phi    
+ N^2 f(\phi)\,  {^{(3)}}R \Big)  -  \dot \phi^2 - 2 N^2  V(\phi)=0  \, ,
\ea
and,
\ba
\label{pert-eq2}
\big( 2 H f +  f_{,\phi} \dot \phi \big) \partial^2 N_1 - 2 f \partial^2 \dot \zeta =0\, .
\ea
As expected, these equations are algebraic in the sense that $\dot N_1$ and  $\dot \psi$ do not appear in these equations. 
Solving the second equation with the appropriate boundary conditions at spatial infinity  yields,
\ba
\label{N1-sol}
N_1= \frac{ \dot \zeta}{H  \cal{F}} \, ,\quad \quad   {\cal F}\equiv 1+ \frac{f_{,\phi} \dot \phi}{2 H f} \, .
\ea
This equation should be compared with the corresponding equation in 
\cite{Maldacena:2002vr} where $f(\phi)=1$ and ${\cal F}=1$.

Using the value of $N_1$  from Eq. (\ref{N1-sol}) in the first constraint 
equation (\ref{pert-eq1}), yields,
\ba
\label{psi_chi_G}
\psi= -\frac{\zeta}{a^2 H {\cal F} } + \chi \, , \quad \quad 
\partial^2 \chi =\frac{\dot \phi^2}{2 H^2}  \frac{{\cal G}}{{\cal F}^2}\, \dot \zeta \, ,
\quad \quad
{\cal G}\equiv 1+ \frac{3 M_{\rm P}^2 f_{,\phi}^2}{2f } \, .
\ea
Again, this equation should be compared with the corresponding equation in  \cite{Maldacena:2002vr} where $f(\phi)=1$ and ${\cal G}=1$. 
In conclusion, compared to the analysis of  \cite{Maldacena:2002vr}, we have two new parameters ${\cal F}$ and ${\cal G}$ because $f(\phi)$ is time-varying during inflation. This is a direct effect of working in the Jordan frame, where the effective gravitational coupling is time-dependent.

After obtaining the solutions for $N$ and $N^i$, we can plug their values into the action and obtain the quadratic action in terms of $\zeta$. After using various integrations by parts, the quadratic action for the scalar perturbations takes the following form, 
\begin{align}\label{actionper}
	S^{(2)}_{\zeta} = \frac{1}{2}\int \dd\tau \dd^3x ~ z_s^2 \Big(
	\zeta'^2 - (\partial_i \zeta)^2
	\Big) \, ,
\end{align}
in which a prime in this section 
denotes the derivative with respect to the conformal time $\dd \tau = \dd t/a$ and
\begin{align}\label{zss}
z_s^2 &\equiv  a^2 \frac{\dot{\phi}^2}{H^2}  \frac{{\cal G}}{{\cal F}^2} \, .
\end{align}
In the limit where $f(\phi)=1$ (i.e. $\xi=0$) we obtain ${\cal G}= {\cal F}=1$, and  \eqref{zss} coincides with that of \cite{Maldacena:2002vr}. Also note that the action (\ref{actionper}) describes perturbations with a sound speed $c_s=1$. 

We comment that the scalar perturbations with a non-minimal coupling 
 in the Jordan frame were also 
studied in \cite{Motohashi:2019tyj}, who obtained a different result for the action compared to our result. In particular, they obtained sound speed $c_s \neq 1$ for scalar perturbation and their normalization factor $z_s$ is different from ours. It is possible that these mismatches originated from discarding the total derivative term in Eq. (\ref{GC}) in the analysis of \cite{Motohashi:2019tyj} as we cautioned above. 

Before finding the solution for the scalar modes in Fourier space $\zeta_k$, it is worth investigating the behaviour of $\zeta$ as it leaves the horizon during inflation. Since the gradient term is negligible outside the horizon, we can safely ignore the last term in Eq. \eqref{actionper}, obtaining
\begin{align}
	\partial_\tau \left( z_s^2 \zeta_k' \right) \simeq 0 \, ,
\end{align}
with the following solution,
\begin{align}
	\label{zeta_outside}
	\zeta_{k}(\tau) \simeq D_{k} + C_k \int_{\tau_*}^{\tau} \frac{\dd \tau'}{z_s(\tau')^2} \, ,
\end{align}
in which $D_k, C_k$ are constants and $\tau_*$ indicates the moment of horizon crossing. As seen from the definition of $z_s^2$ in Eq. \eqref{zss} and the functions $\cal F$ and $\cal G$, $z_s^2$ scales like $a^2$ while $\cal F$ and $\cal G$ evolve slowly. As a result, the integral term in 
Eq. \eqref{zeta_outside} represents the decaying solution of $\zeta_k$, and the dominant solution is the constant part. Correspondingly, as in standard single-field scenarios, the curvature perturbation is frozen on superhorizon scales.

Now, let us decompose the perturbations in the Fourier space, 
\begin{equation}\label{Fourier}
\zeta(\tau,\textbf{x})=\int{\frac{d^{3}k}{(2\pi)^{3}}\zeta_{k}(\tau)e^{i\textbf{k.x}}} \, .
\end{equation}
Defining the canonical field, 
\begin{equation}\label{nu}
v_{k}\equiv z_{s} \zeta_{k} \, ,
\end{equation}
and using Eqs. (\ref{nu}) and (\ref{Fourier}) in Eq. (\ref{actionper}), the quadratic action in Fourier space becomes,
\begin{equation}
	S^{(2)}_{\zeta} = \frac{1}{2}\int \dd\tau \dd^3k ~ \bigg[ |v^{\prime}_{k}|^{2}-\Big(k^{2}-\frac{z_{s}^{\prime\prime}}{z_{s}} \Big)|v_{k}|^{2}\bigg].
\end{equation}
Varying the above action with respect to $v_k$ gives the corresponding Mukhanov-Sasaki equation, 
\begin{equation}\label{zsprime}
v^{\prime\prime}_{k}+\Big( k^{2}- \frac{z_{s}^{\prime\prime}}{z_{s}}\Big)v_{k}=0.
\end{equation}
Using the definition of $z_{s}$ in Eq. \eqref{zss} and  the standard relation,
\ba
a(\tau) \simeq -\frac{1}{(1- \epsilon) H \tau},
\ea
with some effort, one can show that, 
\begin{align}\label{fraczs}
\frac{z_{s}^{\prime\prime}}{z_{s}}\simeq \frac{1}{\tau^{2}}\bigg[2+9\epsilon\big(1+\frac{2}{9}\beta\big)-3\eta_{_V}-12\xi \big(1+2Q-\frac{\eta_{_V}}{4}\frac{\phi^{2}}{M_{P}^{2}}\big)
+\mathcal{O}(\epsilon^{2}, \xi^{2}, \epsilon \xi) \bigg],
\end{align}
where $\beta$ and $Q$ are defined in Eqs. \eqref{beta} and \eqref{epsilon_H} respectively, and $\eta_{_V}=M_{P}^{2} V_{,\phi\phi}/V$ is the second SR parameter in terms of the scalar potential. As expected, Eq. (\ref{fraczs}) shows that the non-minimal coupling changes the effective mass of the scalar fluctuations. 

Recasting \eqref{fraczs} into the following form \cite{Bassett:2005xm},
\begin{align}\label{fraczs1}
\frac{z_{s}^{\prime\prime}}{z_{s}}=\frac{\nu_{s}^{2}-(1/4)}{\tau^{2}} \, ,
\end{align}
the index $\nu_s$ is given by, 
\begin{align}\label{nus}
\nu_{s}\simeq \frac{3}{2}+3\epsilon\big(1+\frac{2}{9}\beta\big)-\eta_{V}-4\xi \Big(1+2Q-\frac{\eta_{V}}{4}\frac{\phi^{2}}{M_{P}^{2}}\Big)
+\mathcal{O}(\epsilon^{2}, \xi^{2}, \epsilon \xi) \,.
\end{align}

Finally, using Eq. (\ref{fraczs1})  in Eq. (\ref{zsprime}) and imposing the Bunch-Davies (Minkowski) initial condition for the modes deep inside the horizon, the mode function of curvature perturbation 
 is obtained to be, 
\begin{align}
\zeta_{k} = \frac{\sqrt{\pi |\tau|}}{2 z_{s}} e^{i(1+2\nu_{s})\frac{\pi}{4}} H_{\nu_{s}}^{(1)}(-k\tau) \, ,
\end{align}
in which $H_{\nu_{s}}^{(1)}$ is the Hankel function of the first kind. This solution has the standard form, and the effects of the non-minimal coupling $\xi$ are captured in the index $\nu_s$ and the normalization factor $z_s$. 

The power spectrum of $\zeta_{k}$ is defined as usual by,
\begin{align}\label{powers}
P_{\zeta}(k) = \frac{\pi |\tau|}{4z_{s}^{2}} \big|H_{\nu_{s}}^{(1)}(-k\tau)\big|^2 .
\end{align}
At the end of inflation $(k\tau\rightarrow 0)$, the Hankel function is approximated as follows, 
\begin{align}\label{HH}
H_{\nu_{s}}^{(1)}(-k\tau) \rightarrow  -\frac{i}{\pi}\Gamma(\nu_{s})\Big(-\frac{k\tau}{2}\Big)^{-\nu_{s}}.
\end{align}
Correspondingly, the dimensionless power spectrum $\mathcal{P}_s(k)$   on superhorizon scales takes the following form, 
\begin{align}\label{dimenps}
\mathcal{P}_s(k) \equiv \frac{k^3}{2\pi^2}P_{\zeta}(k)= \frac{H^{2}}{8\pi^{2}M_{\rm P}^2\,\epsilon_{\phi}}\frac{\mathcal{F}^2}{\mathcal{G}}(-k\tau)^{3-2\nu_s}\, ,
\end{align}
in which  $\epsilon_{\phi} \equiv \dot{\phi}^2/(2M_{\rm P}^2H^2)$ is the first SR parameter in terms of the inflaton velocity. Note that in the presence of $\xi$, $\epsilon_\phi \neq \epsilon$. The above power spectrum can be equivalently written in the following form, 
\begin{align}\label{dimenps2}
	\mathcal{P}_s(k)= \frac{H^{2}}{8\pi^{2}M_{\rm P}^2\,\epsilon_{\phi}}\frac{\mathcal{F}^2}{\mathcal{G}}(-k_\star\tau)^{n_s-1}\left(\frac{k}{k_{\star}}\right)^{n_s-1} \,;
	\hspace{1cm}
	n_s -1 = 3-2\nu_s
\end{align}
where $n_s$ is the scalar spectral index. We also introduced a reference scale $k_\star$, which is useful when comparing with CMB observations\footnote{In the Planck observation \cite{Planck:2018jri}, for example, the data are normalized with respect to the pivot scale $k_\star \simeq 0.05 \,{\rm Mpc}^{_{-1}}$, while this value for WMAP was $0.002\, \,{\rm Mpc}^{_{-1}}$ \cite{WMAP:2010qai}.}. The power spectrum obtained in Eq. (\ref{dimenps2}) still carries an explicit time dependence through the factor $(-k_\star\tau)^{n_s-1}$. However, the physically relevant amplitude of curvature perturbations is determined at the moment of horizon crossing, defined by $k_{\star} = a_{\star} H_{\star}$.  At this moment, the mode exits the Hubble radius and subsequently freezes on super-horizon scales as discussed around Eq. \eqref{zeta_outside}. Using the relation  $\tau_{\star} \simeq -1/(a_{\star}H_{\star})$,  one finds  $-k_\star\tau_{\star} \simeq 1 $, which removes the residual time dependence. This procedure ensures that the spectrum is evaluated at the time when each mode becomes classical, and its amplitude remains conserved thereafter.

The above results were general, valid for any potential. Now, in order to compare with cosmological observations, we consider monomial inflationary potentials. This allows parameters such as the spectral index to be expressed directly in terms of the SR parameters, the CR index, and the non-minimal coupling. For the monomial potentials  $V \propto \phi^n$, using  Eq. \eqref{nus} and the background solution Eq. \eqref{phi-sol-SR}, the spectral index is obtained to be, 
\begin{align}
	\label{ns_phistar-phic}
	n_s \simeq 1- 6\epsilon + 2\eta_{_V} + \frac{16\xi}{f(\phi_\star)} + 2\xi(4-n-2\epsilon)(n-1) +\mathcal{O}(\epsilon^{2}, \xi^{2}, \epsilon \xi) \,.
\end{align}
The first three terms above are the contributions as in standard minimal coupling models \cite{Bassett:2005xm}, while the last two terms are the modifications from the non-minimal coupling. In Sec. \ref{sec:CMB}, we investigate the effects of the non-minimal coupling on the CMB observables.
\subsection{Tensor Perturbations}
 Now let us consider the tensor perturbations. Since the non-minimal coupling affects the gravitational coupling, i.e., $f(\phi)\neq 1$, we expect it to modify the tensor perturbations as well. 
 
Constructing the quadratic action associated with the tensor perturbations, we obtain, 
\ba
S^{(2)}_\gamma = \frac{M_{\rm P}^2}{8}\int \dd\tau \dd^3x ~ z_t^2 \Big[
	(\gamma'_{ij})^2 - (\partial_l \gamma_{ij})^2 \Big] \, ,
\ea
in which we have defined, 
\ba
\label{zt}
z_t^2 \equiv a^2 f(\phi) \, .
\ea
Note the effect of the non-minimal coupling $f(\phi)$ which appears in the normalization factor $z_t$. 

In the  Fourier space the mode function is expanded as follows, 
\ba
	h_{ij}(\tau, {\bf x})  = \int \frac{\dd^3 k}{(2\pi)^3} \sum_{\lambda=+,\times} \epsilon^\lambda_{ij}(k) ~h^\lambda_{\bf k}(\tau) ~e^{i {\bf k} \cdot {\bf x}}\, ,
\ea
in which  the polarization vectors  satisfy the following relations,
\ba
	\epsilon_{ii} = k^i \epsilon_{ij} = 0 \, , \quad  \quad 
	\epsilon^s_{ij}(k) ~ \epsilon^{s'}_{ij}(k) = 2 \delta_{s s'} \, .
\ea

Defining the canonically normalized field via $u_\bfk^\lambda \equiv z_t \, h_{\bfk}^\lambda$, the corresponding mode equation is given by, 
\ba
\label{u-eq}
{u^\lambda_k}'' + \left( k^2 - \frac{z_t''}{z_t}\right) u^\lambda_k = 0 \, .
\ea
To obtain the mode function of tensor perturbations, we need to calculate the quantity $z_t''/z_t$, which plays the role of the effective mass of tensor perturbations. 
From Eq. (\ref{zt}) we obtain,
\ba
\frac{z_t''}{z_t} =  \frac{a''}{a}+ \frac{f''}{2f} - \frac{f'^2}{4 f^2} + \frac{a'}{a} \frac{f'}{f} \, .
\label{zppOz}
\ea
The first term above is given by 
\ba
\label{appOa}
\frac{a''}{a}= \frac{2 + 3 \epsilon}{\tau^2} + {\cal O} (\epsilon^2) \, ,
\ea
while  from  $f(\phi)$  given in Eq. \eqref{f} the remaining new contributions are obtained to be,
\begin{align}
	\label{f_Y}
	\frac{f''}{2f} - \frac{f'^2}{4 f^2} + \frac{a'}{a} \frac{f'}{f}
	&= \frac{f_{, \phi}}{2 f} \phi'' + \big( \frac{f_{, \phi \phi}}{2 f} - \frac{f_{, \phi}^2}{4 f^2}   \big) \phi'^2 -\frac{1}{\tau}  \frac{f_{, \phi}}{ f} \phi'
	\nonumber\\
	&\equiv \frac{3 Y}{\tau^2}\, .
\end{align}
Now, we use the background solution for 
$\phi(N)$ to calculate the new variable $Y$. If the mode of interest $k_\star$ leaves the horizon during the SR regime, we use the background solution given in Eq. (\ref{phi-sol-SR}), and obtain
\ba
\label{Y-n}
Y_\star &\simeq&  \xi \frac{ n\phic^2+(n-4)\phi_\star^2}{\phi_c^2 + \phi_\star^2} + {\cal O}(\epsilon^2, \xi^2, \epsilon \xi)
\,; 
\hspace{1.5cm}
\text{(SR phase)}
\ea
in which $Y_\star = Y|_{\phi = \phi_\star}$ indicates the value of $Y$  when  the mode $k_\star$ leaves the horizon. From the above expression for $Y$ we realize that $Y<0$ for $\xi<0$ and $n>2$.

As the field is slowly rolling, the quantity $Y(\tau)$ runs as well. 
More specifically, using the formula Eq. \eqref{epsilon_H} for $\epsilon$, we obtain,
\ba
\label{Y-ep_n}
Y_\star \simeq \frac{-2\epsilon}{n\Big[1+\left(\frac{\phic}{\phi_\star}\right)^2\Big] -2}
\,; 
\hspace{1.5cm}
\text{(SR phase).}
\ea
This equation shows that $Y_\star>-\epsilon$ for $n\gtrsim 2$,
while it can have noticeable running during inflation.  

If the mode $k_\star$ leaves the horizon close to the CR regime, by using \eqref{phi_CR}, we obtain
\begin{align}
	\label{Y_CR}
	Y_\star \simeq  \frac{n}{12}\beta(\beta+3) \Big(
	\frac{\phi_\star}{\phicr}-1
	\Big)+ {\cal O}(\phi_\star-\phicr)^2
	\hspace{1.5cm}
	\text{(CR phase)} \, .
\end{align}

After substituting Eq. \eqref{f_Y} into Eq. \eqref{zppOz}, one can define the index $\nu_t$ for the tensor perturbations via,
\ba
\frac{z_t''}{z_t} \simeq \frac{2+ 3 \epsilon + 3 Y_\star}{\tau^2}
 \equiv \frac{\nu_t^2 -\frac{1}{4}}{\tau^2} \, ,
\ea
and then obtain,
\ba
\label{nut}
\nu_t = \frac{3}{2} + \epsilon +  Y_\star \, .
\ea
Similar to the scalar perturbations, the index $\nu_t$ characterizes the scale dependence of tensor perturbations. The deviation of  $\nu_t$ from the value $3/2$ generates a tensor spectral tilt. 

Imposing the Bunch-Davies initial condition for the modes deep inside the horizon, the solution of the tensor mode function is obtained to be, 
\ba
\label{ut-sol}
u_k^\lambda(\tau) = \frac{1}{2} \sqrt{- \pi \tau} H_{\nu_t}^{(1)} (- k \tau) \, .
\ea
Having obtained the mode function, we can calculate the tensor power spectrum. Taking into account the normalization prefactor and noting that we have two independent polarizations,  we obtain
\ba
\calP_\gamma = \frac{k^3}{2 \pi^2} \Big( 2\times \frac{4}{z_t^2} \  | u_k^\lambda|^2 \Big) \, ,
\ea
which on superhorizon limit $k \tau \rightarrow 0$, yields,
\ba
\label{P-gamma}
\calP_\gamma (k) = \frac{2 H^2}{ \pi^2 M_{\rm P}^2 f(\phi)} 
 (-k\tau) ^{3- 2 \nu_t} \, .
\ea
Considering the pivot scale $k_\star$, we obtain
\begin{align}
	\calP_\gamma (k) = \frac{2 H^2}{ \pi^2 M_{\rm P}^2 f(\phi)} (-k_\star\tau)^{n_t}\left(\frac{k}{k_{\star}}\right)^{n_t} \,;
\end{align}
in which the tilt of tensor perturbations is given by
\ba
\label{nt}
n_t = 3-2\nu_t
= -2 \epsilon - 2 Y_\star \, ,
\ea
where we have used \eqref{nut}. 

As expected, the effective gravitational coupling with $f(\phi) \neq 1$ appears in the amplitude of tensor perturbations. In addition,  we see that the tensor tilt $n_t$ is shifted by an amount $-2 Y_\star$. Considering Eq. \eqref{Y-ep_n} during the SR regime, one can show that the tilt $n_t$ is shifted by a positive amount for $n \gtrsim 1$.  On the other hand, for the modes leaving the horizon during the CR regime,  one obtains a negative shift for  $n_t$. This can be seen using 
Eq. \eqref{Y_CR} along with Fig. \ref{fig:V_eff}. 

Now let us look at the tensor-to-scalar ratio $r$ defined via,  
\begin{align}\label{rnew}
	r \equiv \frac{\calP_\gamma (k)}{\mathcal{P}_s(k)} \, .
\end{align} 
Since we assumed $f(\phi)>1$ in Eq. \eqref{f} with $\xi<0$, Eq. \eqref{P-gamma} shows that the amplitude of the tensor perturbations is reduced by a factor $f(\phi)^{-1}$ compared to GR. Correspondingly,  using Eqs. (\ref{dimenps}) and (\ref{P-gamma}), one obtains,
\begin{align}\label{rnew}
r \simeq 16 \epsilon_\phi\Big(
 \frac{{\cal G}}{\, {\cal F}^2 \, f}
\Big)\Big|_{\phi=\phi_\star}
\end{align}
where the functions ${\cal F}$ and ${\cal G}$ are defined in Eqs. \eqref{N1-sol} and \eqref{psi_chi_G}, respectively. For a monomial potential $V(\phi) \propto \phi^n$, the above relation is simplified to, 
\begin{align}
	\label{r_monomial}
r\simeq 
\frac{8 M_{\rm P}^2 n^2}{\phi_\star^2} + 8n\left(8-n
\right)\xi + {\cal O}(\epsilon^2, \epsilon \xi, \xi^2)
\end{align}
The first term above is the standard formula for monomial potential in models with a minimal coupling, $r \simeq 16 \epsilon_{_V}$ where $\epsilon_{_V} \equiv \frac{M_{\rm P}^2}{2}\left(
V_{,\phi}/V \right)^2$. As seen,  $\xi <0$ leads to a negative shift in $r$ for $n<8$. This plays an important role when comparing the models involving non-minimal coupling with CMB observations studied in the next section.


\section{CMB Constraints}
\label{sec:CMB}
In this section, we look at the predictions of our model for the CMB observables. The main interests are in the two observables $(n_s, r)$ which are constrained in various CMB observations.  Recently, these observables have been constrained by the Planck, BICEP/Keck (BK), and ACT datasets. These constraints put strong bounds on different inflationary scenarios in the $(n_{s}, r)$ plane, and some of these scenarios have been ruled out or disfavoured, including the original Starobinsky $R^2$ inflation model.
As we reviewed in section \ref{sec:intro}, considering models with the non-minimal coupling can improve the agreement between the monomial inflationary potentials and observational data \cite{Kallosh:2025rni, Wang:2025dbj, Yuennan:2025mlg, Yuennan:2026fcn, Ahmed:2025rrg, Gao:2025onc, Gonuguntla:2026rkw}.

Having the analytical formula for the spectral index \eqref{ns_phistar-phic} and the tensor-to-scalar ratio \eqref{r_monomial} for the monomial potential $V \propto \phi^n$, one can look for the predictions of the model in the  $(n_s, r)$ plane for various values of $\xi$ and $n$. To do this procedure analytically,  one solves for $\phi_\star(\Nf)$  and then eliminates $\phi_\star(\Nf)$ to obtain  $n_s(n, \xi, \Nf )$ and $r(n, \xi, \Nf )$. 

For the monomial potential $V(\phi)\propto \phi^n$, the observables  $n_s$ and $r$ in the minimal regime ($\xi=0$)  are given by \cite{Weinberg:2008zzc},
\begin{align}
	n_s \Big|_{_{\xi =0}} &\simeq 1 -\frac{2 (n +2)}{n +4 \Nf} \,,
	\\
	r\Big|_{_{\xi =0}} &\simeq \frac{16 n }{n +4 {\Nf}} \,.
\end{align}
In the limit where $|\xi| \ll 1$, from Eqs. \eqref{ns_phistar-phic} and \eqref{r_monomial}, the corrections in these quantities induced 
from the non-minimal coupling 
are given by, \begin{align}
	n_s &= n_s \Big|_{_{\xi =0}} + \Delta n_s(\xi) \,,
	\\
	r &= r\Big|_{_{\xi =0}} + \Delta r(\xi) \,,
\end{align} 
in which,
\begin{align}
	\Delta n_s(\xi) &\simeq \frac{16\xi}{f(\phi_\star)} + 2\xi(4-n-2\epsilon)(n-1) +\mathcal{O}(\epsilon^{2}, \xi^{2}, \epsilon \xi) \,,
	\label{nsnew1}
	\\
	\Delta r(\xi) &\simeq 8n\left(8-n
	\right)\xi + {\cal O}(\epsilon^2, \epsilon \xi, \xi^2) \,.
	\label{rnew1}
\end{align}
Hence, the positive (negative) value of $\{\Delta n_s, \Delta r\}$ increases (decreases) the value of $\{n_{s}, r\}$.

These results are useful to understand the effects of $\xi$ on the $(n_s, r)$ parameter space. Considering the negative conformal coupling $(\xi<0)$ in 
Eq. (\ref{rnew1}), the correction to $r$ is negative for $n<8$.  As a result, it causes a reduction in the overall value of $r$ compared to the minimal models. Since many monomial inflationary scenarios predict the amount of $r$ that exceeds the recent observational upper bounds,  this negative contribution becomes especially significant. On the other hand, Eq. (\ref{nsnew1}) demonstrates the deviation of the spectral index from its minimal values.  The first term in Eq. (\ref{nsnew1}) is always negative (assuming $\xi<0$) while the sign of the second term depends on $n$. For $1 \le n < 4$, the second term is also negative, so $n_s$ is smaller compared to the minimal models. However,  for $n<1$ and $n\ge 4$, the second term is positive, so it can compensate for the negative contribution from the first term
such that the total result in $\Delta n_s$ is a positive shift.

\begin{figure}[t!]
	\begin{center}
		\includegraphics[scale=0.3]{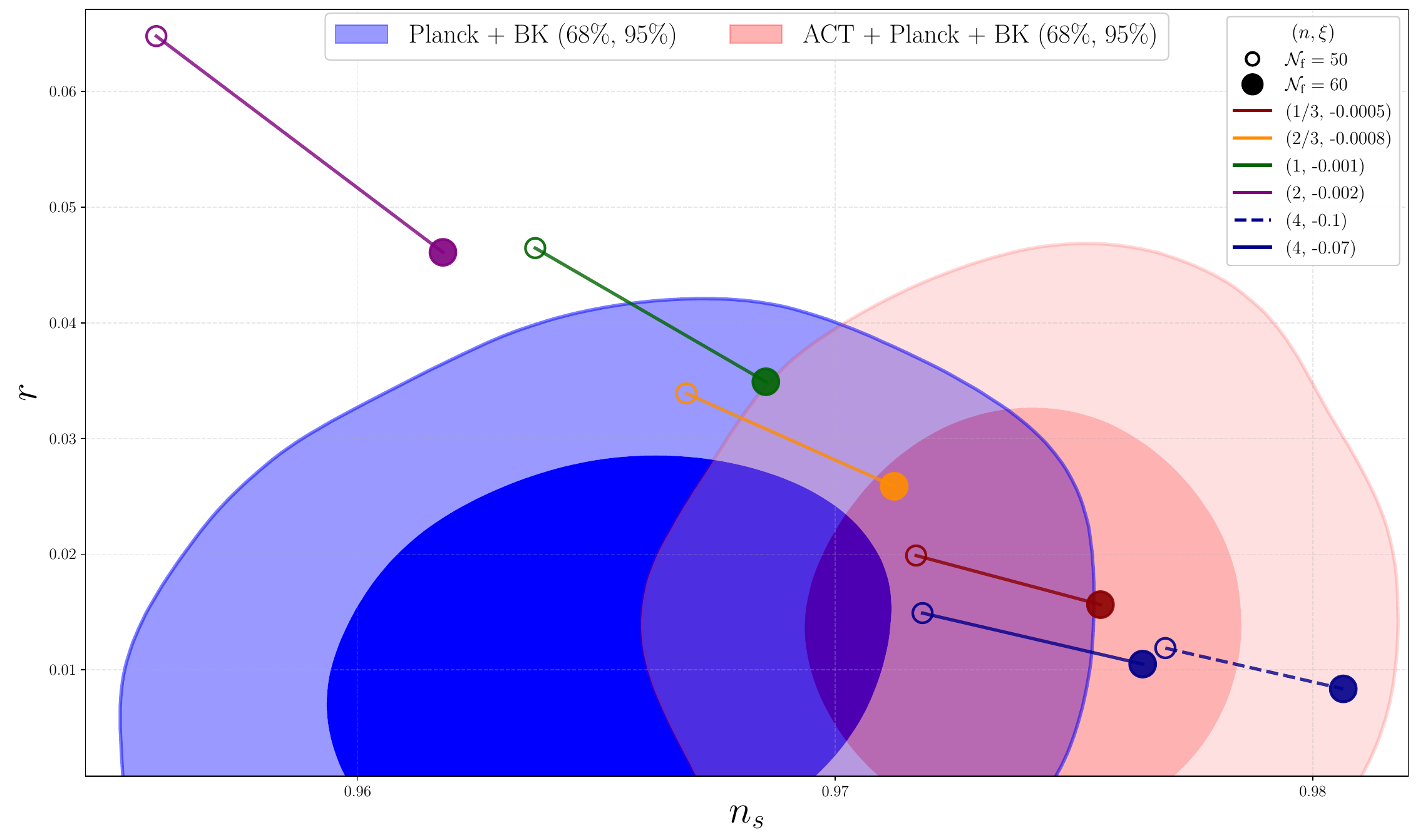}
	\end{center}
	\caption{Constraints in the $(n_s, r)$ plane from the Planck data \cite{Planck:2018vyg} (blue regions) and the ACT+Planck \cite{AtacamaCosmologyTelescope:2025nti} (pink regions) for various values of the model parameters $(n, \xi)$. The plot includes $68\%$ and $95\%$ confidence contours. The filled and empty circles represent the inflationary predictions with $\Nf=60$ and $\Nf=50$, respectively. As seen, the quartic model $n=4$ with $\xi \simeq -0.1$ is consistent with the recent data. Although the non-minimal coupling $\xi$ lowers $r$ for the quadratic ($n=2$) and linear models ($n=1$), the models are still disfavoured as $n_s$ moves towards a smaller value than preferred by the data. } 
	\label{ns-rt}
\end{figure}

Fig. \ref{ns-rt} shows the predictions of the monomial inflationary potentials for different values of non-minimal coupling $(\xi<0)$ in the $(n_s,r)$ plane for $\Nf=50-60$ in connection with the observational data. Different lines represent different values of the model parameters $(n, \xi)$. As shown, in all models with $\xi<0$  the value of $r$ is suppressed compared to the minimal models. Consequently, the points move downward in the $(n_{s}, r)$ plane and come closer to the observationally desired regions requiring small values of $r$. For example, in the quadratic model $n=2$ with $\xi=-0.002$, $r$ is lowered from $0.13$ (when $\xi=0$) to $0.05$ while the spectral index $n_s$ is shifted to smaller values than the case when $\xi=0$. This behaviour is also repeated in the fractional power models such as $n=1/3$ and $n=2/3$ as seen in the plot. More interestingly, the effect of the non-minimal coupling $\xi$ for the quartic model $n=4$ is different; as $|\xi|$ becomes larger, $r$ decreases while $n_s$  increases. For 
instance, the quartic model with $\xi \simeq -0.1$  is in good agreement with the recent ACT data.
This result is consistent with the conclusion in the Planck collaboration that the quartic potential $V(\phi)\propto\phi^4$, which is ruled out at high statistical significance for a minimally coupled scalar field, can be reconciled with the Planck and BICEP/Keck data. In their convention, where the non-minimal coupling appears as $\frac{1}{2}\xi_{\rm P}\phi^2 R$, they obtained a $95\%$ CL lower limit $\log_{10}\xi_{\rm P} > -1.5$~\cite{Planck:2018jri}. Translating to our convention with $f(\phi)=1-\xi(\phi/M_{\rm P})^2$,  this constraint becomes $\xi < -10^{-1.5} \approx -0.0316$.

\begin{figure}[t]
	\begin{center}
		\includegraphics[scale=0.35]{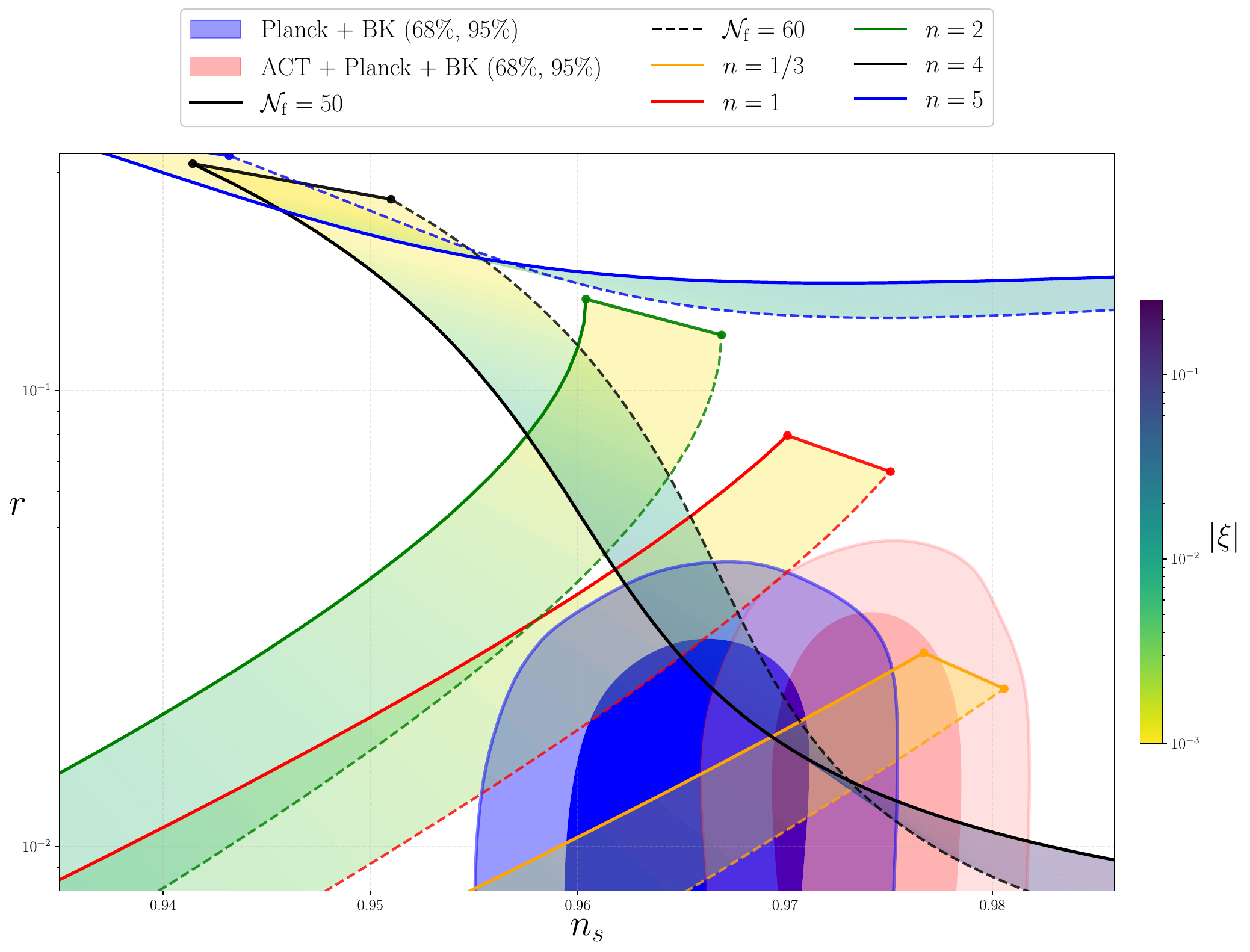}
	\end{center}
	\caption{$(n_s, r)$  predicted by the non-minimal coupling inflation model for various values of $n$ while $\xi$ is varied continuously.  The contours are the $1\sigma$ and $2\sigma$ observational constraints. 
The coloured trajectories denote different values of $n$, and the points along each curve show varying $\xi <0$. The minimal coupling model ($\xi=0$) is located at the upper end of each trajectory with the largest value of $r$.   Increasing  $|\xi|$ for  $n=4$ lowers  $r$ while at the same time increasing $n_s$  such that the predictions of the model shift toward the regions where ACT+Planck+BICEP/Keck observations have overlaps.}
	\label{Delta_ns_r}
\end{figure}

Fig. \ref{Delta_ns_r} illustrates the behaviour of monomial inflation for minimal coupling $(\xi=0)$ and non-minimal coupling $(\xi<0)$ for different values of $n$ and $\Nf= 50-60$ while $\xi$ is varied continuously. In this Figure, the coloured trajectories denote different values of $n$, and the points along each curve indicate varying values of $|\xi|$. As seen, for small values of $n$, the results move far from the observationally preferred region. Among the cases studied, the quartic model $n=4$ with conformal coupling $\xi \sim -0.1$ is in good agreement with the observational data. 
The conclusion is that a negative $\xi$ lowers $r$, while depending on $n$, the spectral index may move towards the left or right compared to the minimal model. 

\section{Conclusions and Remarks}
\label{sec:conclusions}

In this work, we studied models of single-field inflation with a non-minimal coupling $-\xi \phi^{2} R$. This study is well-motivated both theoretically and observationally. On the theoretical side, a non-minimal coupling is a natural contribution in the theory involving GR and a scalar field, as it is vastly studied in the context of quantum field theory in curved spacetime. A non-minimal coupling of the form $-\xi \phi^{2} R$ is a renormalizable term which is expected to appear in the low-energy corrections to the underlying theory. 
Furthermore, in the special case of quartic potential $\lambda \phi^4$, the non-minimal coupling is understood as a conformal coupling in which the theory is classically conformal invariant when $\xi=1/6$. On the other hand, on the observational side,  models with a non-minimal coupling can better fit the cosmological observations, such as the ACT data, which prefer a small tensor-to-scalar ratio and a somewhat larger value of the spectral index $n_s$ compared to the original Planck data.

We have performed the analysis entirely in the Jordan frame. This is in contrast with the usual approach in which one goes to the Einstein frame after a conformal transformation in which the effects of the non-minimal coupling is absorbed into a redefinition of the inflaton field and the modification of the potential. While both methods yield the same physical results, we have argued that working in the Jordan frame keeps the underlying physics and mathematical analysis more transparent. More specifically, the non-minimal coupling $-\xi \phi^{2} R$ appears as a new contribution to the effective mass. Correspondingly, one can easily track the competition between this 
induced mass and the original potential $V(\phi)$ defined in the theory. 
A negative value of $\xi$ causes the inflaton to slow down, yielding a longer period of SR inflation with the same initial value of the inflaton field compared to the models with the minimal coupling.  As a result, even the monomial potential  $V\propto \phi^n$ with $n=2, 4$ does not need a large super-Planckian initial field value to obtain the desired number of e-folds. A direct implication is that with $\xi<0$ one obtains a lower value of $r$ in the $(n_s, r)$ parameter space. Furthermore, with appropriate balance between the driving force $V_{, \phi}$ and the friction term $\xi \phi$, one can have a period of CR inflation. These effects can be easily followed in the Jordan frame where the effective potential is the sum of the monomial potential and the induced mass term $-\xi \phi^{2} R \simeq -12 H^2 \phi^2$.

When studying perturbations in the Jordan frame, special attention must be paid to a total derivative term in the $3+1$ decomposition of the  Ricci scalar. Including this contribution, we computed the quadratic actions for the scalar and tensor perturbations and derived analytical results for  $n_s$ and  $r$, which were then tested against the latest Planck and ACT observations. Our analysis reveals that $\xi<0$ consistently lowers $r$, whereas its effect on $n_s$ is controlled by the power $n$. For $n\geq4$, the shift in $n_s$ differs qualitatively from the cases of $n<4$. These features allow us to fully understand the behaviour of the monomial potential $V \propto \phi^n$ in the $(n_s, r)$ plane.  A key result was that the quartic model is in good agreement with both Planck and the recent ACT data.

The current study can be extended in various directions. A natural extension would be to consider non-monomial potentials such as the symmetry-breaking (i.e., Higgs-like potential) and $\alpha$-attractor models and perform the analysis in the Jordan frame. Another direction to proceed is to study non-Gaussianity in the setups with a non-minimal coupling. As we have seen, in the presence of a non-minimal coupling, the system can enter a period of CR inflation in which it is expected that a non-negligible level of non-Gaussianity to be generated. Roughly speaking, we expect that the amplitude of non-Gaussianity to scale like $f_{\rm NL} \sim \xi$. This is an interesting effect which requires careful investigation.

\vspace{1cm}

{\bf Acknowledgments:} The work of H. F. is supported by the INSF of Iran under the grant number 4045105.





\bibliography{references}

\end{document}